\documentclass[doublecol]{epl2} 

\usepackage{amsmath}
\usepackage{amsfonts}
\usepackage{amssymb}
\usepackage{graphics,graphicx}

\title{Black brane in asymptotically Lifshitz spacetime and viscosity/entropy ratios in Horndeski gravity}
\shorttitle{Black brane in asymptotically Lifshitz spacetime and viscosity/entropy ratios in Horndeski gravity} 

\author{F. A. Brito\inst{1,2} \and F. F. Santos\inst{2}}
\shortauthor{F. A. Brito and F. F. Santos}
\institute{                    
\inst{1} Departamento de F\'\i sica, Universidade Federal de Campina
Grande, Caixa Postal 10071, 58109-970, Campina Grande PB, Brazil\\
\inst{2} Departamento de F\'\i sica, Universidade Federal da Para\'iba, Caixa Postal 5008, 58051-970, Jo\~ao Pessoa PB, Brazil}
	

\abstract{We investigate black brane solutions in asymptotically Lifshitz spacetime in 3+1-dimensional Horndeski gravity, {which admit a critical exponent fixed at $z=1/2$. The cosmological constant depends on $z$ as $\Lambda=-(1+2z)/L^{2}$}. We compute the shear viscosity in the 2+1-dimensional dual boundary field theory via holographic correspondence. We investigate the violation of the bound for viscosity to entropy density ratio of $\eta/s\geq1/(4\pi)$ at $z=1/2$.}

\begin{document}

\maketitle

\section{1. Introduction}
The celebrated Einstein gravity has been supported by strong observational evidence in many astrophysical scenarios. Despite this, there are still fundamental problems such as dark matter, dark energy and inflationary phase of the early Universe to be well-understood in this framework. One of the principal attempts to deal with such problems in Einstein gravity concerns to coupling the theory to scalar fields. Such efforts have led to the development of the now well-known Galileon theories which are scalar-tensor theories \cite{Nicolis:2008in}. Particularly, these studies have led to the rediscovery of {Horndeski's gravity that is the most} general scalar-tensor theory that was originally discovered in 1974 \cite{Horndeski:1974wa,Charmousis:2011bf,Charmousis:2011ea,Starobinsky:2016kua,Bruneton:2012zk,Anabalon:2013oea,Cisterna:2014nua,Heisenberg:2018vsk}. It is characterized by a single scalar-tensor theory with second-order field equations and second-order energy-momentum tensor. The Lagrangian producing second-order equations of motion as discussed in \cite{Deffayet:2011gz,Anabalon:2013oea,VanAcoleyen:2011mj,Gomes:2015dhl,Rinaldi:2016oqp,Cisterna:2017jmv,Gleyzes:2013ooa,Zumalacarregui:2013pma,Cisterna:2014nua} includes four arbitrary functions of the scalar field and its kinetic term. The term we shall focus our attention on includes a nonminimal coupling between the standard scalar kinetic term and the Einstein tensor. Besides the cosmological interest, this theory has also {attracted attention in astrophysics.  In a more recent investigation, it was shown to admit the construction of black holes} that develop Hawking-Page phase transitions at a  critical temperature \cite{Anabalon:2013oea,Rinaldi:2012vy}. Other examples of spherically symmetric solutions in Horndeski theory in the context of the solar system and further astrophysical scenarios can also be found \cite{Cisterna:2015yla,Cisterna:2016vdx,Brihaye:2016lin}, for instance, in the study of perihelion shift and light bending \cite{Bhattacharya:2016naa} and in the issues involving properties of spinning gyroscope and the Gravity Probe B experiment \cite{Mukherjee:2017fqz}.

The AdS/CFT correspondence (holographic description) \cite{Maldacena:1997re,Gubser:1998bc,Witten:1998qj,Aharony:1999ti} has become an important tool to explore strongly coupled field theories. Black branes are fundamental as gravity duals in this correspondence. They, for instance, can be dual to strongly coupled plasma and allow to determine hydrodynamic and thermodynamic properties. {These objects have been proposed recently \cite{Cisterna:2017jmv}  in the context of the Horndeski gravity in a particular sector of the theory}, well known as the K-essence sector, which is supported by axion scalar fields that depend on the horizon coordinates. Further scenarios involving descriptions of static black branes supported by axionic scalars/two 3-form fields were proposed in \cite{Bardoux:2012aw}. These solutions have planar (or toroidal) horizons. 

In quantum critical systems the Lifshitz scaling $t\rightarrow \lambda^{z}t$, $x_{i}\rightarrow\lambda x_{i}$ develops, where $z$ stands for a critical exponent. This scaling is similar to the scaling invariance of the pure AdS spacetime ($z=1$) in Poincaré coordinates. The holographic point of view suggests that such scaling realizes an isometry in the spacetime metric as long as the radial coordinate scales as $r\rightarrow\lambda^{-1}r$ \cite{Bertoldi:2009dt,Bertoldi:2010ca,Bertoldi:2009vn}. Holographic description of models involving Lifshitz scaling has attracted strong interest in recent years, mainly due to applications in condensed matter systems \cite{Kachru:2008yh}. Especially, planar black holes involving Lifshitz superconductors with an axion field as proposed in \cite{Tallarita:2014bga} are of special interest in the context of the AdS/CFT correspondence due to the application in non-conventional superconductor systems.

More recently, {several holographic issues} in the context of the Horndeski gravity have been put forward in \cite{Jiang:2017imk,Baggioli:2017ojd,Liu:2018hzo,Li:2018kqp,Li:2018rgn}. {Among other quantities}, an important relation well known as the shear viscosity to entropy density ratio \cite{Feng:2015oea,Policastro:2001yc,Policastro:2002se,Kovtun:2004de,Sadeghi:2018vrf,Kovtun:2003wp} can be computed in the dual conformal field theory. {This ratio}, which is given by the bound $\eta/s\geq1/(4\pi)$, can be violated by the addition of higher-order curvature terms \cite{Kats:2007mq,Brigante:2007nu}. However, as shown in \cite{Feng:2015oea,Feng:2015wvb} by using the Horndeski gravity this ratio can be simply violated by {adjusting parameters}, without the presence of any such curvature term in the bulk. In our analyzes the violation of the `universal' Kovtun-Son-Starinet (KSS) \cite{Kovtun:2003wp,Policastro:2001yc} bound $\eta/s=1/(4\pi)$ is due to the critical exponent {$z=1/2$ and sufficiently large ratio of parameters $|\alpha/\gamma\Lambda|$}, which implies $\eta/s<1/(4\pi)$ and confirms the aforementioned violations.

In the present study, we find black brane solutions in asymptotically Lifshitz spacetime in Horndeski gravity. Lifshitz black holes with a time-dependent scalar field in this theory were studied in \cite{Bravo-Gaete:2013dca}. {Here, we obtain} static Lifshitz black holes for Horndeski parameters related to the cosmological constant $\Lambda=-(1+2z)/L^{2}$ at a fixed critical exponent $z=1/2$. For Einstein gravity, Lifshitz solutions are obtained due to massive vector fields, scalar and/or massless vector fields \cite{Hoyos:2013cba}.  

Finally, to compute the shear viscosity to entropy density living in a 2+1-dimensional dual boundary field theory we consider the black brane background as the gravity dual {assuming that it is extended} along of the two spatial dimensions where the linearization of the field equations is performed. The graviton fluctuations decouple from all modes \cite{DeWolfe:1999cp,Brito:2018pwe} and can be treated separately \cite{Kovtun:2004de}. The shear viscosity is then computed by finding the retarded Green's function through the holographic correspondence \cite{Maldacena:1997re,Gubser:1998bc,Witten:1998qj,Aharony:1999ti,Policastro:2001yc,Hartnoll:2016tri,Lucas:2015vna,Cremonini:2018jrx}. 

This work is summarized as follows. In Sec.~{\bf2} we address the issue of finding black hole solutions in asymptotically Lifshitz spacetime. In Sec.~{\bf3} we compute the shear viscosity/entropy density ratio \cite{Policastro:2001yc,Policastro:2002se,Hartnoll:2016tri,Eling:2011ms,Eling:2011ct,Lucas:2015vna,Son:2002sd,Jain:2015txa} and we show a violation of the `universal' bound $\eta/s=1/(4\pi)$. Finally, in Sec.~{\bf4} we present our conclusions.

\section{2. Black brane solutions in asymptotically Lifshitz spacetime in Horndeski gravity}\label{v0}

In this section we address the issue of finding black brane solutions in asymptotically Lifshitz spacetime \cite{Bertoldi:2009dt,Bertoldi:2010ca} in Horndeski gravity \cite{Horndeski:1974wa,Cisterna:2014nua}. Lifshitz black holes with a time-dependent scalar field in Horndeski's theory have been previously studied in \cite{Bravo-Gaete:2013dca} and black brane solutions in Horndeski gravity have been considered in \cite{Cisterna:2017jmv}. 
The Horndeski Lagrangian is given by 
\begin{eqnarray}
&&\mathcal{L}_{H}=\mathcal{L}_{2}+\mathcal{L}_{3}+\mathcal{L}_{4}+\mathcal{L}_{5},\\
&&\mathcal{L}_{2}=G_{2}(X,\phi),\\
&&\mathcal{L}_{3}=-G_{3}(X,\phi)\Box\phi,\\
&&\mathcal{L}_{4}=G_{4}(X,\phi)R+\partial_{X}G_{4}(X,\phi)\delta^{\mu\nu}_{\alpha\beta}\nabla^{\alpha}_{\mu}\phi\nabla^{\beta}_{\nu}\phi,\\
&&\mathcal{L}_{5}=G_{5}(X,\Phi)G_{\mu\nu}\nabla^{\mu}\nabla^{\nu}\phi\nonumber\\
&&-\frac{1}{6}\partial_{X}G_{5}(X,\phi)\delta^{\mu\nu\rho}_{\alpha\beta\gamma}\nabla^{\alpha}_{\mu}\phi\nabla^{\beta}_{\nu}\phi\nabla^{\gamma}_{\rho}\phi,
\end{eqnarray}
where $X\equiv -\frac{1}{2}\nabla_{\mu}\phi\nabla^{\nu}\phi$. There are special subclasses of this theory \cite{Charmousis:2011bf,Charmousis:2011ea,Starobinsky:2016kua} that constrain the coefficients $G_{k}(X,\phi)$. We shall consider the following interesting subclass with non-minimal kinetic coupling given by the action 
  
\begin{eqnarray}
&&I[g_{\mu\nu},\phi]=\int{\sqrt{-g}d^{4}x\mathcal{L}},\label{10}\\
&&\mathcal{L}=\kappa(R-2\Lambda)-\frac{1}{2}(\alpha g_{\mu\nu}-\gamma G_{\mu\nu})\nabla^{\mu}\phi\nabla^{\nu}\phi.\nonumber
\end{eqnarray}
Note that we have a non-minimal scalar-tensor coupling where we can define a new field $\phi^{'}\equiv\psi$. This field has a dimension of $(mass)^{2}$ and the parameters $\alpha$ and $\gamma$ control the strength of the kinetic couplings, $\alpha$ is dimensionless and $\gamma$ has a dimension of $(mass)^{-2}$. {Thus, the Einstein-Horndeski field equations can be formally written as in the usual way}
\begin{equation}
G_{\mu\nu}+\Lambda g_{\mu\nu}=\frac{1}{2\kappa}T_{\mu\nu},\label{11}
\end{equation}
where $T_{\mu\nu}=\alpha T^{(1)}_{\mu\nu}+\gamma T^{(2)}_{\mu\nu}$ with $\kappa=(16\pi G)^{-1}$ and the scalar field equation is given by 
\begin{equation}
\nabla_{\mu}[(\alpha g^{\mu\nu}-\gamma G^{\mu\nu})\nabla_{\nu}\phi]=0.\label{12}
\end{equation}
The energy-momentum tensors $T^{(1)}_{\mu\nu}$ and $T^{(2)}_{\mu\nu}$ take the following form
\begin{eqnarray}
&&T^{(1)}_{\mu\nu}=\nabla_{\mu}\phi\nabla_{\nu}\phi-\frac{1}{2}g_{\mu\nu}\nabla_{\lambda}\phi\nabla^{\lambda}\phi\nonumber\\
&&T^{(2)}_{\mu\nu}=\frac{1}{2}\nabla_{\mu}\phi\nabla_{\nu}\phi R-2\nabla_{\lambda}\phi\nabla_{(\mu}\phi R^{\lambda}_{\nu)}
\nonumber\\
&&-\nabla^{\lambda}\phi\nabla^{\rho}\phi R_{\mu\lambda\nu\rho}\nonumber\\
&&-g_{\mu\nu}\left[-\frac{1}{2}(\nabla^{\lambda}\nabla^{\rho}\phi)(\nabla_{\lambda}\nabla_{\rho}\phi)\right.\nonumber\\
&&\left.+\frac{1}{2}(\Box\phi)^{2}-(\nabla_{\lambda}\phi\nabla_{\rho}\phi)R^{\lambda\rho}\right]\nonumber\\
&&-(\nabla_{\mu}\nabla^{\lambda}\phi)(\nabla_{\nu}\nabla_{\lambda}\phi)+(\nabla_{\mu}\nabla_{\nu}\phi)\Box\phi
\nonumber\\
&&+\frac{1}{2}G_{\mu\nu}(\nabla\phi)^{2}.\label{g}
\end{eqnarray}
In our case for Einstein-Horndeski gravity, we consider the following {\sl Ansatz} for a general four-dimensional Lifshitz black brane of the form

\begin{equation}
ds^{2}=L^{2}\left(-r^{2z}f(r)dt^{2}+r^{2}d\Omega^{2}_{2,\epsilon}+\frac{dr^{2}}{r^{2}f(r)}\right).\label{me}
\end{equation}
Here $d\Omega^{2}_{2,\epsilon}$ is the metric for the unit $S^{2}$ sphere, plane or hyperboloid corresponding to $\epsilon=1,0,-1$, respectively \cite{Feng:2015oea}. Note we can take $d\Omega^{2}_{2,\epsilon}=\bar{g}_{\mu\nu}dx^{\mu}dx^{\nu}$ to be the metric of constant curvature such that $\bar{R}_{\mu\nu}=\epsilon\bar{g}_{\mu\nu}$. We can also make $d\Omega^{2}_{2,\epsilon}$ as
\begin{equation}
d\Omega^{2}_{2,\epsilon}=\frac{du^{2}}{1-\epsilon u^{2}}+u^{2}d\Omega^{2}_{1},
\end{equation}
where $d\Omega^{2}_{1}$ is the metric of the unit 1-sphere. In \cite{Hui:2012qt} static spherically symmetric configurations of certain Galileons with shift-invariance was first argued to admit a no-hair theorem. The no-hair theorem for Galileons requires that the square of the conserved current $J^{\mu}=(\alpha g^{\mu\nu}-\gamma G^{\mu\nu})\nabla_{\nu}\phi$, defined in (\ref{12}), should not diverge at horizon. Thus, to escape from the no-hair theorem \cite{Hui:2012qt,Bravo-Gaete:2013dca,Rinaldi:2012vy,Babichev:2013cya}, we need to impose that the radial component of the conserved current vanishes identically without restricting the radial dependence of the scalar field \cite{Bravo-Gaete:2013dca}:
\begin{equation}
\alpha g_{rr}-\gamma G_{rr}=0\label{0a}.
\end{equation}
Recalling that $\phi'(r)\equiv \psi(r)$ we can easily note that this condition annihilates $\psi^2(r)$ regardless of its behavior at the horizon. {The metric function $f(r)$ can be found by using the equation (\ref{0a}). Thus, one can show that the equation (\ref{12}) is satisfied by the following solution 

\begin{eqnarray}
f(r)&=&\frac{\alpha L^{2}}{\gamma(2z+1)}-\left(\frac{r_{0}}{r}\right)^{2z+1},\label{scalar.1}\\
\psi^{2}(r)&=&-\frac{2L^{2}\kappa(\alpha+\gamma\Lambda)}{\alpha\gamma r^{2}f(r)},\label{scalar.2}
\end{eqnarray}
{where the scalar field is real for a suitable relation between the parameters --- see its rewritten form below.
Finally, the Einstein-Horndeski field equation (\ref{11})  is satisfied by these equations as long as $z=1$ or $z=1/2$ with {$\epsilon=0$ (the planar case)}}. The former solution corresponds to black brane solution for asymptotically AdS$_4$ spacetime \cite{Bertoldi:2009dt}. We shall focus on the latter case \cite{Tallarita:2014bga,Feng:2015oea}. Note that when $\Lambda=-(1+2z)/L^{2}$  that is in accord with \cite{Bertoldi:2009dt,Bertoldi:2010ca,Bertoldi:2009vn} we have the black hole solution in asymptotically Lifshitz spacetime
\begin{eqnarray}
f(r)&=&-\frac{\alpha}{\gamma\Lambda}-\left(\frac{r_{0}}{r}\right)^{2z+1},\label{il}\\
\psi^{2}(r)&=&\frac{2\kappa(1+2z)(\alpha+\gamma\Lambda)}{\alpha\gamma\Lambda r^{2}f(r)},\label{3w}
\end{eqnarray}
for $z=1/2$. The parameters are defined in the range $-\infty<\alpha/\gamma\Lambda\leq-1$, with $\alpha,\gamma<0$, or $-1\leq\alpha/\gamma\Lambda<0$, with $\alpha,\gamma>0$.

We can still make the following transformation in the metric (\ref{me}) with solution (\ref{il}) 
\begin{eqnarray}
&&f(r)\to-\frac{\alpha}{\gamma\Lambda}{f}(r),r_0^{2z+1}\to-\frac{\alpha}{\gamma\Lambda}\;r_0^{2z+1},\nonumber\\
&&L\to\left(-\frac{{\alpha}}{\gamma\Lambda}\right)^{1/2}L,t\to-\frac{\gamma\Lambda}{\alpha}t,\nonumber\\
&&x\to\left(-\frac{{\gamma\Lambda}}{\alpha}\right)^{1/2}x,y\to\left(-\frac{{\gamma\Lambda}}{\alpha}\right)^{1/2}y,
\end{eqnarray}
in order to put the black hole solution in the standard form
\begin{eqnarray}
{f}(r)&=&1-\left(\frac{r_{0}}{r}\right)^{2z+1},\label{il2}\\
\psi^{2}(r)&=&-\frac{2\kappa(1+2z)(\alpha+\gamma\Lambda)}{\alpha^2 r^{2}{f}(r)}.\label{3w2}
\end{eqnarray}
In addition, the fact that $\psi^2(r\to\infty)=0$ into the action (\ref{10}) ensures that this is a genuine vacuum solution. Note also that from equations (\ref{il2})-(\ref{3w2}) the black hole geometry is regular everywhere (except at the central singularity), the scalar field derivative $\psi(r)$ diverges at horizon \cite{Anabalon:2013oea,Feng:2015oea,Babichev:2013cya}, but the scalar field does not explode at horizon since it approaches to a constant near the horizon as
$\phi^{2}(r)\sim\left((\Lambda L^2(\alpha+\gamma\Lambda)/\alpha^{2}r^{2}_{0}f^{'}(r_{0}))(r-r_{0})\right)+const.$
These facts are in complete agreement with the no-hair theorem as previously discussed and evade the issues raised in \cite{Babichev:2013cya}. The scalar field equation (\ref{3w2}) is a real function outside the horizon since for $r>r_{0}$ we have $f(r>r_{0})>0$, and {the scalar field is real, for example, in the interval $-1<\alpha/\gamma\Lambda<0$, with $\alpha, \gamma>0$}. We can see that at infinity the scalar field itself diverges as $\phi(r)\sim \ln{r}$, but not its derivatives $\psi$ that are the ones present in the action, which are finite at asymptotic infinity \cite{Babichev:2013cya}.
 
In the following, we address the issue of singularity through the curvature invariant given by the Ricci scalar 

\begin{eqnarray}
&&R=\frac{r^{2}f^{''}(r)+3zrf^{'}(r)+2z^{2}f(r)}{L^{2}}\nonumber\\
&&+\frac{5rf^{'}(r)+4zf(r)+6f(r)}{L^{2}},\label{Rt}\\
&&={-\frac{1}{L^{2}}\left[2z^{2}+4z+6-3(z-1)\left(\frac{r_{0}}{r}\right)^{2z+1}\right]}.
\end{eqnarray}

We can see that for $r=0$ and $z=1/2$ we find a curvature singularity since the Ricci scalar diverges, which implies that we found a solution that is in fact a black brane solution. However, for $z=1$, we have $R=-12/L^{2}$, that is, $R=4\Lambda$ which characterizes the AdS$_{4}$ spacetime \cite{Anabalon:2013oea}. 

\section{3. Viscosity/entropy density ratio}\label{v2}

In this section, we present the computation of the shear viscosity in the boundary field theory through holographic correspondence \cite{Feng:2015oea,Sadeghi:2018vrf,Kovtun:2003wp}. We do this in the Horndeski gravity context \cite{Feng:2015oea,Liu:2016njg}, where a black brane solution is found in the presence of asymptotically Lifshitz spacetime. In the gravity side, this planar black brane plays the role of the gravitational dual of a certain fluid. To compute the shear viscosity through holographic correspondence,  we need to linearize the field equations \cite{Kovtun:2004de,Sadeghi:2018vrf,Kovtun:2003wp}. Thus, the effective hydrodynamics in the boundary field theory is constructed in terms of conserved currents and energy-momentum tensor by considering small fluctuations around the black brane background $g_{xy}\rightarrow g_{xy}+h_{xy}$, where $h_{xy}=h_{xy}(t,\vec{x},r)$ \cite{Sadeghi:2018vrf,Kovtun:2003wp,Liu:2016njg,HosseiniMansoori:2018gdu,Baier:2007ix} is a small perturbation --- related issues in braneworlds at Einstein and Horndeski gravity can be seen, e.g., in \cite{DeWolfe:1999cp,Brito:2018pwe}. For the metric (\ref{me}) {in the planar case, i.e., $\epsilon=0$}, we find the fluctuations of Ricci tensor in the form 
\begin{eqnarray}
&&\delta^{(1)}R_{xy}=-\frac{r^{2}f(r)}{2L^{2}}h^{''}_{xy}+\frac{\ddot{h}_{xy}}{2r^{2z}f(r)L^{2}}\nonumber\\
&&-\frac{(rf(r)(z+3)+r^{2}f^{'}(r))}{2L^{2}}h^{'}_{xy}.
\end{eqnarray}
Here we have disregarded the dependence of $h_{xy}$ on $\vec{x}$. Recalling that $\bar{T}_{xy}=T_{xy}-g_{xy}T^{\alpha}_{\alpha}/3$ and using Einstein-Horndeski equation in the Ricci form then $\delta^{(1)}R_{xy}=\delta^{(1)}\bar{T}_{xy}/(2\kappa)$, such that we find a Klein-Gordon-like equation with a position dependent mass as follows
\begin{eqnarray}
\frac{1}{\sqrt{-g}}\partial_{\alpha}(\sqrt{-g}g^{\alpha\beta}\partial_{\beta}h_{xy})=-m^{2}(r)h_{xy}\;,\label{yj}
\end{eqnarray}
where
{
\begin{eqnarray}
&&m^{2}(r)=\frac{2(1+2z)}{3L^{2}}\left[\frac{(4+7z+z^{2})}{(-\alpha(1+2z)/(\gamma\Lambda)+3+6z)}\right]\nonumber\\
&&+\frac{2(1+z-2z^{2})}{3L^{2}(-\alpha(1+2z)/(\gamma\Lambda)+3+6z)}\left(\frac{r_{0}}{r}\right)^{2z+1}\label{jo}.
\end{eqnarray}}
Now considering the following {\sl Ansatz}
\begin{eqnarray}
h_{xy}(x,r)=\int{\frac{d^3k}{(2\pi)^3}}e^{ikx}\chi(r,k),
\end{eqnarray}
with $x=(t,\vec{x}\,)$ and $k=(\omega,\vec{q}\,)$. The equation of motion for the fluctuations assumes the following form
\begin{eqnarray}
\frac{1}{\sqrt{-g}}\partial_{r}(\sqrt{-g}g^{rr}\partial_{r}\chi(r,k))=-m^{2}(r)\chi(r,k)\label{8i}.
\end{eqnarray}
In addition to the mass term, in general, this equation also contains the contribution $k^2=q^2-\omega^2$, but we have considered the limit $\omega\to0$ and spatial momentum $q=0$. {After these considerations, we shall simply consider $\chi(r,k)\equiv\chi(r)$}.

In our present scenario, we analyze viscosity/entropy density ratio for the critical exponent $z=1/2$. Non integer critical exponents have appeared in previous studies, e.g., \cite{Bravo-Gaete:2013dca,MohammadiMozaffar:2017nri,MohammadiMozaffar:2017chk,MohammadiMozaffar:2018vmk,Passos:2016bbc,Anacleto:2018wlj} --- and references therein. Especially in \cite{MohammadiMozaffar:2017nri,MohammadiMozaffar:2017chk} the authors argue that although the Lifshitz critical exponents in the action of a quantum field theory developing Lifshitz symmetry are assumed to be an integer, there is no such limitation indeed since the obtained dispersion relation in the Hamiltonian density associated with the quantized theory shows the exact analytic continuation to noninteger values of $z$. Furthermore,  in \cite{MohammadiMozaffar:2018vmk} the authors also consider noninteger $z>1$ in the problem of entanglement propagation in Lifshitz-type scalar theories.

The general explicit solution for the differential equation (\ref{8i}) can be given by \cite{Liu:2016njg}

\begin{eqnarray}
&&\chi(r)=C_{1}\frac{F_{1}\left(a_{-},a_{+},b_{-};\left(\frac{r_{0}}{r}\right)^{2z+1}\right)}{(r/r_{0})^{\beta_{-}}}\nonumber\\
&&+C_{2}(r/r_{0})^{\beta_{+}}F_{1}\left(d_{-},d_{+},b_{+};\left(\frac{r_{0}}{r}\right)^{2z+1}\right),\label{sl}\\
&&\beta_{\mp}=\pm\frac{\mp\sqrt{9\gamma\Lambda-3\alpha}z\mp 2\sqrt{9\gamma\Lambda-3\alpha}}{2\sqrt{9\gamma\Lambda-3\alpha}}\\
&&\pm\frac{\sqrt{z^{2}\gamma\Lambda-20z\gamma\Lambda-3z^{2}\alpha+4\gamma\Lambda-12z\alpha-12\alpha}}{2\sqrt{9\gamma\Lambda-3\alpha}},\nonumber\\
&&a_{\mp}=-\frac{\mp\sqrt{9\gamma z\Lambda-17\gamma\Lambda-3\alpha z+3\alpha}\sqrt{z-1}}{2(2z+1)\sqrt{9\gamma\Lambda-3\alpha}}-\nonumber\\
&&-\frac{-2z\sqrt{9\gamma\Lambda-3\alpha}-\sqrt{9\gamma\Lambda-3\alpha}}{2(2z+1)\sqrt{9\gamma\Lambda-3\alpha}}-\\
&&-\frac{\sqrt{z^{2}\gamma\Lambda-20z\gamma\Lambda-3z^{2}\alpha+4\gamma\Lambda-12z\alpha-12\alpha}}{2(2z+1)\sqrt{9\gamma\Lambda-3\alpha}}\nonumber\\
&&b_{\mp}=\mp\frac{\mp\sqrt{9\gamma\Lambda-3\alpha}z\mp\sqrt{9\gamma\Lambda-3\alpha}}{(2z+1)\sqrt{9\gamma\Lambda-3\alpha}}+\\
&&+\frac{\sqrt{z^{2}\gamma\Lambda-20z\gamma\Lambda-3z^{2}\alpha+4\gamma\Lambda-12z\alpha-12\alpha}}{(2z+1)\sqrt{9\gamma\Lambda-3\alpha}},\nonumber\\ 
&&d_{\mp}=\frac{\mp\sqrt{9\gamma z\Lambda-17\gamma\Lambda-3\alpha z+3\alpha}\sqrt{z-1}}{2(2z+1)\sqrt{9\gamma\Lambda-3\alpha}}+\nonumber\\
&&+\frac{2z\sqrt{9\gamma\Lambda-3\alpha}+\sqrt{9\gamma\Lambda-3\alpha}}{2(2z+1)\sqrt{9\gamma\Lambda-3\alpha}}+\nonumber\\
&&\frac{\sqrt{z^{2}\gamma\Lambda-20z\gamma\Lambda-3z^{2}\alpha+4\gamma\Lambda-12z\alpha-12\alpha}}{2(2z+1)\sqrt{9\gamma\Lambda-3\alpha}}.
\end{eqnarray} 
On the horizon $r=r_{0}$, {we can see that $\chi$ in the equation (\ref{sl}) diverges for general values of $C_{1}$ and $C_{2}$, because the argument $r_0/r$ of the hypergeometric functions becomes equal to unity. In order to remove this divergence, we first rewrite the equation (\ref{sl}) in terms of gamma functions, such that on the horizon reads}
\begin{eqnarray}
&&\chi(r_{0})=\frac{C_{1}\Gamma(b_{-})\Gamma(a_{-}+a_{+}-b_{-})}{\Gamma(b_{-}-a_{-})\Gamma(b_{-}-a_{+})}\nonumber\\
&&+\frac{C_{2}\Gamma(b_{+})\Gamma(d_{+}+d_{-}-b_{+})}{\Gamma(b_{+}-d_{+})\Gamma(b_{+}-d_{-})}.
\end{eqnarray}
{We can now remove the divergence by choosing}
\begin{eqnarray}
\frac{C_{2}}{C_{1}}=\frac{\Gamma(b_{-})\Gamma(b_{+}-d_{+})\Gamma(b_{+}-d_{-})}{\Gamma(b_{+})\Gamma(b_{-}-a_{-})\Gamma(b_{-}-a_{+})}.
\end{eqnarray}
This allows us to find the following regularized solution
\begin{eqnarray}\label{eq-chi-r0}
\chi(r_{0})=\frac{2\pi C_{1}\Gamma(b_{-})}{\Gamma(b_{-}-a_{-})\Gamma(b_{-}-a_{+})},
\end{eqnarray}
that will be crucial to our following considerations. The integration constant $C_1$ can be set to unity without loss of generality. {To obtain the final result in the form (\ref{eq-chi-r0}) we have made use of the parameters $a_\pm$, $b_\pm$ and $d_\pm$ definitions given above and gamma function properties.}

Let us now focus on the computation of the shear viscosity of the fluid in 2+1-dimensional dual boundary field theory by proceeding as follows. We can see  that we can propose an effective action for the equation (\ref{8i}) written as 
\begin{eqnarray}
&&S=-\frac{1}{16\pi G}\int{\frac{d^{3}kdr}{(2\pi)^{3}}\left(\frac{N(r)}{2}\frac{d\chi(r,k)}{dr}\frac{d\chi(r,-k)}{dr}\right)}\nonumber\\
&&-\frac{1}{16\pi G}\int{\frac{d^{3}kdr}{(2\pi)^{3}}\left(-\frac{M(r)}{2}\chi(r,k)\chi(r,-k)\right)},
\end{eqnarray}
where $N(r)=\sqrt{-g}g^{rr}$ and $M(r)=\sqrt{-g}m^{2}(r)$. This action evaluated on-shell reduces to the surface term
\begin{eqnarray}\label{S_cl}
S=-\left.\frac{1}{16\pi G}\int{\frac{d^{3}k}{(2\pi)^{3}}\left(\frac12{N(r)}\chi(r,k){\partial_r\chi(r,-k)}\right)}\right|_{r_0}^{\infty}.
\end{eqnarray}
From the holographic correspondence \cite{Maldacena:1997re,Gubser:1998bc,Witten:1998qj,Aharony:1999ti,Hartnoll:2016tri,Lucas:2015vna,Son:2002sd,Jain:2015txa,Pang:2009wa} the Green's retarded function can be computed via two-point function from the generator of connected correlation functions on the boundary which is given in terms of the classical action (\ref{S_cl}).

Thus, as is well known, we have that the retarded Green's function reads
\begin{eqnarray}\label{G_Rfinal}
G^{R}_{xy,xy}(\omega,0)=\frac{-2}{16\pi G}(\sqrt{-g}g^{rr}|_{r_{0}})\chi(r,-\omega)\partial_{r}\chi(r,\omega)|_{r_{0}},
\end{eqnarray}
{where we have admitted spatial momentum $q=0$. Since the imaginary part of the Green's function does not dependent on the radial coordinate, we have conveniently chosen to compute it at horizon $r=r_0$}. 

Now, from regularity at the horizon, the derivative of $\chi$ at the horizon is given in terms of $\chi(r_0)$ at leading order in $\omega$ \cite{Chakrabarti:2010xy}. As such, we can write the equation (\ref{G_Rfinal}) as follows 
\begin{eqnarray}
G^{R}_{xy,xy}(\omega,0)=-\frac{2i\omega L^{2}r^{2}_{0}}{16\pi G}\chi^{2}(r_{0}).
\end{eqnarray}
The shear viscosity \cite{Policastro:2001yc,Policastro:2002se,Hartnoll:2016tri,Jain:2015txa,Son:2002sd} is then given by
\begin{eqnarray}
\eta&=&-\lim_{\omega\rightarrow 0}\frac{1}{2\omega}{\rm Im}G^{R}_{xy,xy}\\
&=&\frac{L^{2}r^{2}_{0}}{16\pi G}\chi^{2}(r_{0}),\label{sh-to-s}
\end{eqnarray}
where $s=r^{2}_{0}L^{2}/4G$ is the entropy density \cite{Tallarita:2014bga}. Thus, we can now write the shear viscosity/entropy density ratio as in the form

\begin{eqnarray}
\frac{\eta}{s}=\frac{1}{4\pi}\left[\frac{2\pi\Gamma(b_{-})}{\Gamma(b_{-}-a_{-})\Gamma(b_{-}-a_{+})}\right]^{2}.\label{g0u}
\end{eqnarray}

\begin{table}[!ht]
\begin{center}
\begin{tabular}{|c|c|c|c|c|c|c|} \hline
$-\alpha/(\gamma\Lambda)$&$\eta/s$\\ \hline
$5$&$13.14/4\pi>1/4\pi$\\ \hline
$10$&$2.44/4\pi>1/4\pi$\\ \hline
$15$&$0.57/4\pi<1/4\pi$\\ \hline
$40$&$0.053/4\pi<1/4\pi$\\ \hline
\end{tabular}
\end{center}
\caption{The table shows the shear viscosity to entropy density ratio $\eta/s$ for some values of $-\alpha/\gamma\Lambda$ with critical exponent $z=1/2$.}
\label{Tb1}
\end{table}

One should notice that from equation (\ref{sh-to-s}), that is
\begin{eqnarray}
\frac{\eta}{s}=\frac{1}{4\pi}\chi^{2}(r_{0}),\label{sh-to-s-2}
\end{eqnarray}
our analyzes show that for the fixed critical exponent $z=1/2$ the `universal' bound $\eta/s\geq1/(4\pi)$ is not violated, for example, as $-\alpha/(\gamma\Lambda)=5,10$, but it is violated if $|\alpha/\gamma\Lambda|$ is sufficiently large,  for example, as $-\alpha/(\gamma\Lambda)=15,40$ --- see table \ref{Tb1}. Moreover, for the case $\alpha=-\gamma\Lambda$ this bound saturates, i.e., $\eta/s=1/(4\pi)$. This is because in this regime the equation (\ref{3w2}) gives $\psi(r)=0$ which in turns substituting into equation (\ref{8i}) yields $m(r)=0$. Thus, by solving the equation (\ref{8i}), we find $\chi(r)=const.$, everywhere. Then, on the horizon one can assume the solution normalized to unit $\chi(r_0)=1$, which from (\ref{sh-to-s-2}) clearly saturates the bound.

\section{4. Conclusions}\label{v3}

In this paper, we show that black brane with asymptotically Lifshitz spacetime for the critical exponent $z=1/2$ is a solution of Horndeski's theory. From the holographic point of view, the black brane plays the role of the gravitational dual of a boundary quantum field theory that describes a certain viscous fluid. The holographic description has become an important tool to explore strongly coupled field theories, more specifically strongly coupled plasmas. In this sense, the holographic description of hydrodynamics has attracted great interest in the literature for reasons similar to those that have attracted special interest in the application of this correspondence, for example, to non conventional superconductor systems. Especially, in our present system, the holography imposes specific conditions on the parameters. {On the boundary quantum field side, the viscosity/entropy density ratio is $\eta/s<1/(4\pi)$, for example, as $-\alpha/(\gamma\Lambda)=15,40$ with $z=1/2$. In other words, this means that the KSS bound is violated \cite{Foster:2016abe} in our setup in the context of the Horndeski gravity. On the other hand, for example, as $-\alpha/(\gamma\Lambda)=5,10$ with $z=1/2$, we have $\eta/s>1/(4\pi)$. In both cases, the system does not develop an ideal fluid since $\eta$ either increases or decreases with the variation of the parameters but does not vanish. This fact clearly characterizes a viscous fluid with entropy production \cite{Rangamani:2009xk}. The case $\alpha=-\gamma\Lambda$ saturates the bound i.e., $\eta/s=1/(4\pi)$.
Finally, we have confirmed previous conclusions that the `universal' KSS bound can be violated in a wide class of conventional theories with no higher-derivative terms in the Lagrangian.} 

\acknowledgments

We would like to thank CNPq, CAPES, and CNPq/PRONEX/FAPESQ-PB (Grant No. 165/2018) for partial financial support. FAB acknowledges support from CNPq
(Grant no. 312104/2018-9).


\begin{thebibliography}{0}

\bibitem{Nicolis:2008in} 
  A.~Nicolis, R.~Rattazzi and E.~Trincherini,
  {\it The Galileon as a local modification of gravity},
  Phys.\ Rev.\ D {\bf 79}, 064036 (2009),
  [arXiv:hep-th/0811.2197].

\bibitem{Horndeski:1974wa} 
  G.~W.~Horndeski,
  {\it Second-order scalar-tensor field equations in a four-dimensional space},
  Int.\ J.\ Theor.\ Phys.\  {\bf 10}, 363 (1974).

\bibitem{Charmousis:2011bf} 
  C.~Charmousis, E.~J.~Copeland, A.~Padilla and P.~M.~Saffin,
  {\it General second order scalar-tensor theory, self tuning, and the Fab Four},
  Phys.\ Rev.\ Lett.\  {\bf 108}, 051101 (2012),
  [arXiv:1106.2000 [hep-th]].
	
\bibitem{Charmousis:2011ea} 
  C.~Charmousis, E.~J.~Copeland, A.~Padilla and P.~M.~Saffin,
  {\it Self-tuning and the derivation of a class of scalar-tensor theories},
  Phys.\ Rev.\ D {\bf 85}, 104040 (2012),
  [arXiv:1112.4866 [hep-th]].

\bibitem{Starobinsky:2016kua} 
  A.~A.~Starobinsky, S.~V.~Sushkov and M.~S.~Volkov,
  {\it The screening Horndeski cosmologies},
  JCAP {\bf 1606}, 007 (2016),
  [arXiv:1604.06085 [hep-th]].
	
\bibitem{Bruneton:2012zk} 
  J.~P.~Bruneton, M.~Rinaldi, A.~Kanfon, A.~Hees, S.~Schlogel and A.~Fuzfa,
  {\it Fab Four: When John and George play gravitation and cosmology},
  Adv.\ Astron.\  {\bf 2012}, 430694 (2012),
  [arXiv:1203.4446 [gr-qc]].

	\bibitem{Cisterna:2014nua} 
  A.~Cisterna and C.~Erices,
  {\it Asymptotically locally AdS and flat black holes in the presence of an electric field in the Horndeski scenario},
  Phys.\ Rev.\ D {\bf 89}, 084038 (2014),
  [arXiv:gr-qc/1401.4479].
	
		\bibitem{Anabalon:2013oea} 
  A.~Anabalon, A.~Cisterna and J.~Oliva,
  {\it Asymptotically locally AdS and flat black holes in Horndeski theory},
  Phys.\ Rev.\ D {\bf 89}, 084050 (2014),
  [arXiv:gr-qc/1312.3597 [gr-qc]].

	\bibitem{Heisenberg:2018vsk} 
  L.~Heisenberg,
  {\it A systematic approach to generalisations of General Relativity and their cosmological implications},
  arXiv:1807.01725 [gr-qc].
	
		
	\bibitem{VanAcoleyen:2011mj} 
  K.~Van Acoleyen and J.~Van Doorsselaere,
  {\it Galileons from Lovelock actions},
  Phys.\ Rev.\ D {\bf 83}, 084025 (2011),
  [arXiv:1102.0487 [gr-qc]].
	
  \bibitem{Deffayet:2011gz} 
  C.~Deffayet, X.~Gao, D.~A.~Steer and G.~Zahariade,
  {\it From k-essence to generalised Galileons},
  Phys.\ Rev.\ D {\bf 84}, 064039 (2011),
  [arXiv:1103.3260 [hep-th]].
		
	\bibitem{Gomes:2015dhl} 
  A.~R.~Gomes and L.~Amendola,
  {\it The general form of the coupled Horndeski Lagrangian that allows cosmological scaling solutions},
  JCAP {\bf 1602}, 035 (2016),
  [arXiv:1511.01004 [gr-qc]].
	
	\bibitem{Rinaldi:2016oqp} 
  M.~Rinaldi,
  {\it Mimicking dark matter in Horndeski gravity},
  Phys.\ Dark Univ.\  {\bf 16}, 14 (2017),
  [arXiv:1608.03839 [gr-qc]].
			
	\bibitem{Gleyzes:2013ooa} 
  J.~Gleyzes, D.~Langlois, F.~Piazza and F.~Vernizzi,
  {\it Essential Building Blocks of Dark Energy},
  JCAP {\bf 1308}, 025 (2013),
  [arXiv:1304.4840 [hep-th]].
	
	\bibitem{Zumalacarregui:2013pma} 
  M.~Zumalacárregui and J.~García-Bellido,
  {\it Transforming gravity: from derivative couplings to matter to second-order scalar-tensor theories beyond the Horndeski Lagrangian},
  Phys.\ Rev.\ D {\bf 89}, 064046 (2014),
  [arXiv:1308.4685 [gr-qc]].
	
	\bibitem{Cisterna:2017jmv} 
  A.~Cisterna, M.~Hassaine, J.~Oliva and M.~Rinaldi,
  {\it Axionic black branes in the k-essence sector of the Horndeski model},
  Phys.\ Rev.\ D {\bf 96}, no. 12, 124033 (2017),
	[arXiv:1708.07194 [hep-th]].
	
  \bibitem{Rinaldi:2012vy} 
  M.~Rinaldi,
  {\it Black holes with non-minimal derivative coupling},
  Phys.\ Rev.\ D {\bf 86}, 084048 (2012),
  [arXiv:1208.0103 [gr-qc]].
	
\bibitem{Cisterna:2015yla} 
  A.~Cisterna, T.~Delsate and M.~Rinaldi,
  {\it Neutron stars in general second order scalar-tensor theory: The case of nonminimal derivative coupling},
  Phys.\ Rev.\ D {\bf 92}, no. 4, 044050 (2015),
  [arXiv:1504.05189 [gr-qc]].
	
\bibitem{Cisterna:2016vdx} 
  A.~Cisterna, T.~Delsate, L.~Ducobu and M.~Rinaldi,
  {\it Slowly rotating neutron stars in the nonminimal derivative coupling sector of Horndeski gravity},
  Phys.\ Rev.\ D {\bf 93}, no. 8, 084046 (2016),
  [arXiv:1602.06939 [gr-qc]].
	
\bibitem{Brihaye:2016lin} 
  Y.~Brihaye, A.~Cisterna and C.~Erices,
  {\it Boson stars in biscalar extensions of Horndeski gravity},
  Phys.\ Rev.\ D {\bf 93}, no. 12, 124057 (2016),
  [arXiv:1604.02121 [hep-th]].
  
\bibitem{Bhattacharya:2016naa} 
  S.~Bhattacharya and S.~Chakraborty,
  {\it Constraining some Horndeski gravity theories},
  Phys.\ Rev.\ D {\bf 95}, no. 4, 044037 (2017),
  [arXiv:1607.03693 [gr-qc]].
	
\bibitem{Mukherjee:2017fqz} 
  S.~Mukherjee and S.~Chakraborty,
  {\it Horndeski theories confront the Gravity Probe B experiment},
  Phys.\ Rev.\ D {\bf 97}, no. 12, 124007 (2018),
  [arXiv:1712.00562 [gr-qc]].
  	
\bibitem{Maldacena:1997re} 
  J.~M.~Maldacena,
  {\it The Large N limit of superconformal field theories and supergravity},
  Int.\ J.\ Theor.\ Phys.\  {\bf 38}, 1113 (1999)
  [Adv.\ Theor.\ Math.\ Phys.\  {\bf 2}, 231 (1998)],
	[hep-th/9711200].
  
\bibitem{Gubser:1998bc} 
  S.~S.~Gubser, I.~R.~Klebanov and A.~M.~Polyakov,
  {\it Gauge theory correlators from noncritical string theory},
  Phys.\ Lett.\ B {\bf 428}, 105 (1998),
  [hep-th/9802109].
  
\bibitem{Witten:1998qj} 
  E.~Witten,
  {\it Anti-de Sitter space and holography},
  Adv.\ Theor.\ Math.\ Phys.\  {\bf 2}, 253 (1998),
  [hep-th/9802150].
	
\bibitem{Aharony:1999ti} 
  O.~Aharony, S.~S.~Gubser, J.~M.~Maldacena, H.~Ooguri and Y.~Oz,
  {\it Large N field theories, string theory and gravity},
  Phys.\ Rept.\  {\bf 323}, 183 (2000),
  [hep-th/9905111].
	
	\bibitem{Bardoux:2012aw} 
  Y.~Bardoux, M.~M.~Caldarelli and C.~Charmousis,
  {\it Shaping black holes with free fields},
  JHEP {\bf 1205}, 054 (2012),
  [arXiv:1202.4458 [hep-th]].
	  
  	\bibitem{Bertoldi:2009dt} 
  G.~Bertoldi, B.~A.~Burrington and A.~W.~Peet,
  {\it Thermodynamics of black branes in asymptotically Lifshitz spacetimes},
  Phys.\ Rev.\ D {\bf 80}, 126004 (2009),
  [arXiv:0907.4755 [hep-th]].
	
	\bibitem{Bertoldi:2010ca} 
  G.~Bertoldi, B.~A.~Burrington and A.~W.~Peet,
  {\it Thermal behavior of charged dilatonic black branes in AdS and UV completions of Lifshitz-like geometries},
  Phys.\ Rev.\ D {\bf 82}, 106013 (2010),
  [arXiv:1007.1464 [hep-th]].
	
  \bibitem{Bertoldi:2009vn} 
  G.~Bertoldi, B.~A.~Burrington and A.~Peet,
  {\it Black Holes in asymptotically Lifshitz spacetimes with arbitrary critical exponent},
  Phys.\ Rev.\ D {\bf 80}, 126003 (2009),
  [arXiv:0905.3183 [hep-th]].
  
	
\bibitem{Kachru:2008yh} 
  S.~Kachru, X.~Liu and M.~Mulligan,
  {\it Gravity duals of Lifshitz-like fixed points},
  Phys.\ Rev.\ D {\bf 78}, 106005 (2008),
  [arXiv:0808.1725 [hep-th]].

	
	\bibitem{Tallarita:2014bga} 
  G.~Tallarita,
  {\it Holographic Lifshitz Superconductors with an Axion Field},
  Phys.\ Rev.\ D {\bf 89}, no. 10, 106005 (2014),
  [arXiv:1402.4691 [hep-th]].
	
	\bibitem{Jiang:2017imk} 
  W.~J.~Jiang, H.~S.~Liu, H.~Lu and C.~N.~Pope,
  {\it DC Conductivities with Momentum Dissipation in Horndeski Theories},
  JHEP {\bf 1707}, 084 (2017),
  [arXiv:1703.00922 [hep-th]].
	
	\bibitem{Baggioli:2017ojd} 
  M.~Baggioli and W.~J.~Li,
  {\it Diffusivities bounds and chaos in holographic Horndeski theories},
  JHEP {\bf 1707}, 055 (2017),
  [arXiv:1705.01766 [hep-th]].
	
\bibitem{Liu:2018hzo} 
H.~S.~Liu,
{\it Violation of Thermal Conductivity Bound in Horndeski Theory},
Phys.\ Rev.\ D {\bf 98}, no. 6, 061902 (2018),
[arXiv:1804.06502 [hep-th]].

\bibitem{Li:2018kqp} 
  Y.~Z.~Li and H.~Lu,
  {\it $a$-theorem for Horndeski gravity at the critical point},
  Phys.\ Rev.\ D {\bf 97}, no. 12, 126008 (2018),
  [arXiv:1803.08088 [hep-th]].

 \bibitem{Li:2018rgn} 
 Y.~Z.~Li, H.~Lu and H.~Y.~Zhang,
 {\it Scale Invariance vs. Conformal Invariance: Holographic Two-Point Functions in Horndeski Gravity},
 arXiv:1812.05123 [hep-th].
	
	\bibitem{Feng:2015oea} 
  X.~H.~Feng, H.~S.~Liu, H.~Lü and C.~N.~Pope,
  {\it Black Hole Entropy and Viscosity Bound in Horndeski Gravity},
  JHEP {\bf 1511}, 176 (2015),
  [arXiv:1509.07142 [hep-th]].
	
	\bibitem{Kovtun:2003wp} 
  P.~Kovtun, D.~T.~Son and A.~O.~Starinets,
  {\it Holography and hydrodynamics: Diffusion on stretched horizons},
  JHEP {\bf 0310}, 064 (2003),
  [hep-th/0309213].
	
	\bibitem{Kovtun:2004de} 
  P.~Kovtun, D.~T.~Son and A.~O.~Starinets,
  {\it Viscosity in strongly interacting quantum field theories from black hole physics},
  Phys.\ Rev.\ Lett.\  {\bf 94}, 111601 (2005),
  [hep-th/0405231].
	
	\bibitem{Sadeghi:2018vrf} 
  M.~Sadeghi,
  {\it Black Brane Solution in Rastall AdS Massive Gravity and Viscosity Bound},
  Mod.\ Phys.\ Lett.\ A {\bf 33}, no. 37, 1850220 (2018),
  [arXiv:1809.08698 [hep-th]].
	
	\bibitem{Policastro:2001yc} 
  G.~Policastro, D.~T.~Son and A.~O.~Starinets,
  {\it The Shear viscosity of strongly coupled N=4 supersymmetric Yang-Mills plasma},
  Phys.\ Rev.\ Lett.\  {\bf 87}, 081601 (2001),
	[hep-th/0104066].
	
	\bibitem{Policastro:2002se} 
  G.~Policastro, D.~T.~Son and A.~O.~Starinets,
  {\it From AdS / CFT correspondence to hydrodynamics},
  JHEP {\bf 0209}, 043 (2002),
  [hep-th/0205052].
	
	\bibitem{Kats:2007mq} 
  Y.~Kats and P.~Petrov,
  {\it Effect of curvature squared corrections in AdS on the viscosity of the dual gauge theory},
  JHEP {\bf 0901}, 044 (2009),
  [arXiv:0712.0743 [hep-th]].
	
	\bibitem{Brigante:2007nu} 
  M.~Brigante, H.~Liu, R.~C.~Myers, S.~Shenker and S.~Yaida,
  {\it Viscosity Bound Violation in Higher Derivative Gravity},
  Phys.\ Rev.\ D {\bf 77}, 126006 (2008),
  [arXiv:0712.0805 [hep-th]].
	
	\bibitem{Feng:2015wvb} 
  X.~H.~Feng, H.~S.~Liu, H.~Lü and C.~N.~Pope,
  {\it Thermodynamics of Charged Black Holes in Einstein-Horndeski-Maxwell Theory},
  Phys.\ Rev.\ D {\bf 93}, no. 4, 044030 (2016),
  [arXiv:1512.02659 [hep-th]].
	
	\bibitem{Bravo-Gaete:2013dca} 
  M.~Bravo-Gaete and M.~Hassaine,
  {\it Lifshitz black holes with a time-dependent scalar field in a Horndeski theory},
  Phys.\ Rev.\ D {\bf 89}, 104028 (2014),
  [arXiv:1312.7736 [hep-th]].

	\bibitem{Hoyos:2013cba} 
  C.~Hoyos, B.~S.~Kim and Y.~Oz,
  {\it Bulk Viscosity in Holographic Lifshitz Hydrodynamics},
  JHEP {\bf 1403}, 050 (2014),
  [arXiv:1312.6380 [hep-th]].
		
  \bibitem{DeWolfe:1999cp} 
  O.~DeWolfe, D.~Z.~Freedman, S.~S.~Gubser and A.~Karch,
  {\it Modeling the fifth-dimension with scalars and gravity},
  Phys.\ Rev.\ D {\bf 62}, 046008 (2000),
  [hep-th/9909134].
	
	\bibitem{Brito:2018pwe} 
  F.~A.~Brito and F.~F.~Dos Santos,
  {\it Braneworlds in Horndeski gravity},
  arXiv:1810.08196 [hep-th].
  	
	\bibitem{Hartnoll:2016tri} 
  S.~A.~Hartnoll, D.~M.~Ramirez and J.~E.~Santos,
  {\it Entropy production, viscosity bounds and bumpy black holes},
  JHEP {\bf 1603}, 170 (2016),
  [arXiv:1601.02757 [hep-th]].
	
 \bibitem{Lucas:2015vna} 
  A.~Lucas,
  {\it Conductivity of a strange metal: from holography to memory functions},
  JHEP {\bf 1503}, 071 (2015),
  [arXiv:1501.05656 [hep-th]].
	
\bibitem{Cremonini:2018jrx} 
  S.~Cremonini, M.~Cvetic and I.~Papadimitriou,
  {\it Thermoelectric DC conductivities in hyperscaling violating Lifshitz theories},
  JHEP {\bf 1804}, 099 (2018),
  [arXiv:1801.04284 [hep-th]].
	
	\bibitem{Eling:2011ms} 
  C.~Eling and Y.~Oz,
  {\it A Novel Formula for Bulk Viscosity from the Null Horizon Focusing Equation},
  JHEP {\bf 1106}, 007 (2011),
  [arXiv:1103.1657 [hep-th]].
	
	\bibitem{Eling:2011ct} 
  C.~Eling and Y.~Oz,
  {\it Holographic Screens and Transport Coefficients in the Fluid/Gravity Correspondence},
  Phys.\ Rev.\ Lett.\  {\bf 107}, 201602 (2011),
  [arXiv:1107.2134 [hep-th]].
	
	
\bibitem{Son:2002sd} 
  D.~T.~Son and A.~O.~Starinets,
  {\it Minkowski space correlators in AdS / CFT correspondence: Recipe and applications},
  JHEP {\bf 0209}, 042 (2002),
  [hep-th/0205051].
	
\bibitem{Jain:2015txa} 
  S.~Jain, R.~Samanta and S.~P.~Trivedi,
  {\it The Shear Viscosity in Anisotropic Phases},
  JHEP {\bf 1510}, 028 (2015),
  [arXiv:1506.01899 [hep-th]].
	
  \bibitem{Hui:2012qt} 
  L.~Hui and A.~Nicolis,
  {\it No-Hair Theorem for the Galileon},
  Phys.\ Rev.\ Lett.\  {\bf 110}, 241104 (2013),
  [arXiv:1202.1296 [hep-th]].
		
  \bibitem{Babichev:2013cya} 
  E.~Babichev and C.~Charmousis,
  {\it Dressing a black hole with a time-dependent Galileon},
  JHEP {\bf 1408}, 106 (2014),
  [arXiv:1312.3204 [gr-qc]].
		

  \bibitem{Liu:2016njg} 
  H.~S.~Liu, H.~Lu and C.~N.~Pope,
  {\it Magnetically-Charged Black Branes and Viscosity/Entropy Ratios},
  JHEP {\bf 1612}, 097 (2016),
  [arXiv:1602.07712 [hep-th]].
  
  \bibitem{HosseiniMansoori:2018gdu} 
  S.~A.~Hosseini Mansoori, V.~Jahnke, M.~M.~Qaemmaqami and Y.~D.~Olivas,
  {\it Holographic complexity of anisotropic black branes},
  arXiv:1808.00067 [hep-th].
		
	\bibitem{Baier:2007ix} 
  R.~Baier, P.~Romatschke, D.~T.~Son, A.~O.~Starinets and M.~A.~Stephanov,
  {\it Relativistic viscous hydrodynamics, conformal invariance, and holography},
  JHEP {\bf 0804}, 100 (2008),
  [arXiv:0712.2451 [hep-th]].
	
	\bibitem{MohammadiMozaffar:2017nri} 
  M.~R.~Mohammadi Mozaffar and A.~Mollabashi,
  {\it Entanglement in Lifshitz-type Quantum Field Theories},
  JHEP {\bf 1707}, 120 (2017),
  [arXiv:1705.00483 [hep-th]].

  \bibitem{MohammadiMozaffar:2017chk} 
  M.~R.~Mohammadi Mozaffar and A.~Mollabashi,
  {\it Logarithmic Negativity in Lifshitz Harmonic Models},
  J.\ Stat.\ Mech.\  {\bf 1805}, no. 5, 053113 (2018),
  [arXiv:1712.03731 [hep-th]].

\bibitem{MohammadiMozaffar:2018vmk} 
  M.~R.~Mohammadi Mozaffar and A.~Mollabashi,
  {\it Entanglement Evolution in Lifshitz-type Scalar Theories},
  JHEP {\bf 1901}, 137 (2019),
  [arXiv:1811.11470 [hep-th]].

\bibitem{Passos:2016bbc} 
  E.~Passos, E.~M.~C.~Abreu, M.~A.~Anacleto, F.~A.~Brito, C.~Wotzasek and C.~A.~D.~Zarro,
  {\it Lifshitz scaling to Lorentz-violating high derivative operator and gamma-ray bursts},
  Phys.\ Rev.\ D {\bf 93}, no. 8, 085022 (2016),
  [arXiv:1603.01558 [hep-th]].

\bibitem{Anacleto:2018wlj} 
  M.~A.~Anacleto, F.~A.~Brito, E.~Maciel, A.~Mohammadi, E.~Passos, W.~O.~Santos and J.~R.~L.~Santos,
  {\it Lorentz-violating dimension-five operator contribution to the black body radiation},
  Phys.\ Lett.\ B {\bf 785}, 191 (2018),
  [arXiv:1806.08273 [hep-th]].
		
	\bibitem{Pang:2009wa} 
  D.~W.~Pang,
  {\it Conductivity and Diffusion Constant in Lifshitz Backgrounds},
  JHEP {\bf 1001}, 120 (2010),
  [arXiv:0912.2403 [hep-th]].
		
	\bibitem{Chakrabarti:2010xy} 
  S.~K.~Chakrabarti, S.~Chakrabortty and S.~Jain,
  {\it Proof of universality of electrical conductivity at finite chemical potential},
  JHEP {\bf 1102}, 073 (2011),
  [arXiv:1011.3499 [hep-th]].
	
	\bibitem{Foster:2016abe} 
  J.~W.~Foster and J.~T.~Liu,
  {\it Spatial Anisotropy in Nonrelativistic Holography},
  arXiv:1612.01557 [hep-th].
		
	\bibitem{Rangamani:2009xk} 
  M.~Rangamani,
  {\it Gravity and Hydrodynamics: Lectures on the fluid-gravity correspondence},
  Class.\ Quant.\ Grav.\  {\bf 26}, 224003 (2009),
  [arXiv:0905.4352 [hep-th]].

\end{thebibliography}
\end{document}